\documentclass{aa}

\usepackage{graphicx}
\usepackage[a]{esvect}
\usepackage{txfonts}

\def\gsim{\mathrel{\raise.5ex\hbox{$>$}\mkern-14mu
             \lower0.6ex\hbox{$\sim$}}}
\def\lsim{\mathrel{\raise.3ex\hbox{$<$}\mkern-14mu
             \lower0.6ex\hbox{$\sim$}}}

\begin{document} 	
	
	\title{High-energy radiation from the pulsar equatorial current sheet}
\titlerunning{ECS without PIC}
\authorrunning{Contopoulos, P\'{e}tri \& Dimitropoulos}

		\author{Ioannis Contopoulos\inst{1*}, J\'{e}r\^{o}me P\'{e}tri\inst{2}
		\and Ioannis Dimitropoulos\inst{1}
	}
		\institute{Research Center for Astronomy and Applied Mathematics, Academy of Athens, Athens 11527, Greece\\
	    	 \inst{*}\email{icontop@academyofathens.gr}
		\and
Universit\'{e} de Strasbourg, CNRS, Observatoire astronomique de Strasbourg, UMR 7550, F-67000 Strasbourg, France}
	
	\abstract
	{Pulsars emit beams of radiation that reveal the extreme physics of neutron star magnetospheres. However, our understanding is incomplete. Recent global particle-in-cell (PIC) simulations have raised several questions that led us to question their validity and their extrapolation to realistic particle Lorentz factors, and electric and magnetic fields.}
	{Our aim was to generate realistic sky maps of high-energy radiation from first principles.}
	{We propose a novel method to study the equatorial current sheet (ECS) where most of the particle acceleration and the high-energy radiation is expected to originate. We first determined its shape and external magnetic field with a steady-state ideal force-free solution. Then, we considered the extra electric and magnetic field components that develop when dissipation is considered. Finally, we studied the particle acceleration and radiation that is due to these extra field components for realistic field and particle parameters.}
	{We generated realistic sky maps of high-energy radiation and compared them with those obtained via PIC simulations. These sky maps may also be closely reproduced using the ECS of the split-monopole solution beyond the light cylinder.}
	{The ECS is probably stabilized by the normal magnetic field component that is due to the global magnetospheric reconnection. Our method helps us better understand the origin of the pulsed high-energy radiation in the pulsar magnetosphere.} 

	
\maketitle

\section{Introduction}
Pulsars, the spinning remnants of massive stars, act as cosmic lighthouses, emitting beams of radiation that reveal the extreme physics of neutron stars. However, our understanding of their magnetospheres—the magnetic environments powering these beams—remains incomplete, thus limiting our understanding of pulsar variability. A reference ideal force-free magnetosphere for an oblique rotator is still lacking.

The most recent global particle-in-cell (PIC) simulations have raised several questions \citep[e.g.,][]{Cerutti2016,Hu2022,Hakobyan2023,Soudais2024,Cerutti2025}. Notably, these simulations yield separatrix regions between open and closed field lines that exhibit a significant thickness beyond the simulation resolution, and a multi-layered internal structure. This is intriguing because the same simulations are able to generate thin current sheets in the equator. Moreover, their closed-line regions terminate well inside the light cylinder, with regions of strong electromagnetic dissipation beyond their tips, something that was not expected by \cite{GJ}. Interestingly, the shape of the tips resembles a pointed Y, thus deviating from the theoretically predicted T shape \citep{Uzdensky2003,Ypoint}.

We do not understand the physical origin of these effects in PIC simulations. Unfortunately, it is in general difficult to differentiate between artificial (numerical) and true (physical) features in the published solutions. The resolution of current global PIC simulations is inadequate  to properly model the microphysics of the equatorial current sheet (ECS). In order to generate pulsar light curves and spectra that may be compared with observations, the simulation results are extrapolated by several orders of magnitude. Unfortunately, there is no agreement among different research groups on the particular method of extrapolation, and as a result, there is still no generally accepted understanding of the physical origin of the high-energy radiation from pulsars.

In this work, we propose a novel way to study the ECS beyond the light cylinder where high-energy radiation is expected to be generated. We assume the shape and internal magnetic field distribution that we obtained with our steady-state ideal force-free physics-informed neural network (PINN) solution for one particular pulsar inclination angle $\lambda=20^\circ$ \citep{Dimitropoulos2026}.\footnote{We have not yet applied our method to higher inclination angles.}  Then, we calculate the extra electric and magnetic field components that develop when dissipation is considered along the ECS. Finally, we investigate particle acceleration with realistic electromagnetic field and particle parameters, and generate realistic sky maps of high-energy radiation. We compare our sky maps with those obtained via PIC simulations, and draw our conclusions about the practical applications of our approach.

\section{The internal structure of the ECS}

\begin{figure}[h!]
		\centering
		\includegraphics[width=1.\columnwidth]{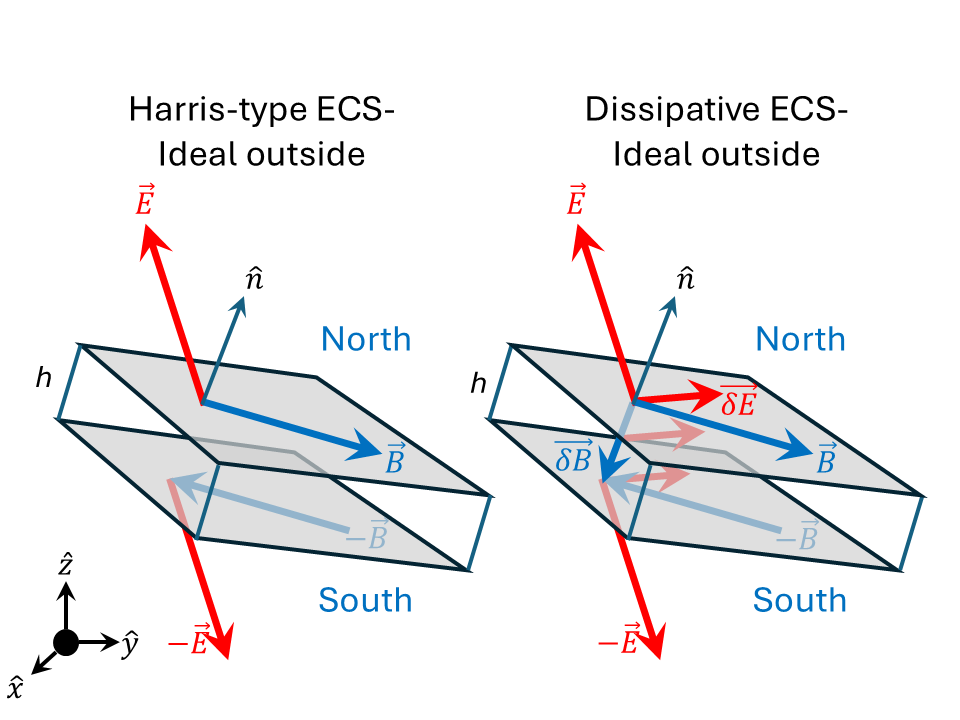}
		\caption{Cross section of the equatorial current sheet (ECS)  along its thickness $h$. Left:  Ideal Harris-type current sheet with perfect field reversal across it and no dissipation. Right: Dissipative current sheet with nonzero extra electric and magnetic field components $\delta E$ and $\delta B$ respectively in its interior. The normal magnetic field component $\delta B$ is absent in local reconnection simulations that start from a Harris-type current sheet. Wherever $\delta E>\delta B$ there is particle acceleration, thus also dissipation in the current sheet. In both cases the ECS lies inside a steady-state corotating ideal force-free  magnetosphere. In the cartoon of the central star in the lower left corner, $z$ is along $\Omega$ and, without loss of generality, the central magnetic dipole $\hat{\mu}$ points in the direction of the north rotation hemisphere, namely $\hat{\mu}\cdot\Omega>0$. The ECS lies beyond the light cylinder at distances  $r\gg h$ from the star.}
		\label{ECS}
\end{figure}

We consider a slice of the ECS as shown in figure~\ref{ECS}. Without loss of generality we  assume that the northern magnetic hemisphere is up, the southern is down, and the angular velocity of rotation vector $\vv{\Omega}$ points up. We  work in a spherical system of coordinates $r,\theta,\phi$ aligned with the direction of rotation. {In the cartesian system shown in figure~\ref{ECS}, the $z$-axis is along $\theta=0$}. We define the unit vector $\hat{n}\equiv (n_r,n_\theta,n_\phi)$ that is normal to the ECS and points in the direction of the northern magnetic hemisphere. 

We  first consider the ideal case without reconnection nor energy dissipation. In that case, the ECS is a perfect Harris-type current sheet along the northern surface of which $\vv{B}\cdot\hat{n}=0$ and
\begin{equation}
\vv{E}=-\hat{\phi}\times \vv{B}\left(\frac{r\ \sin\theta}{R_{\rm lc}}\right)\equiv -\hat{\phi}\times \vv{B}_{\!p}\left(\frac{r\ \sin\theta}{R_{\rm lc}}\right)\equiv \vv{E}_{\!p} ,
\label{North1}
\end{equation}
where the subscript $p$ denotes the poloidal   ($r,\theta$) field component (also called meridional), and $R_{\rm lc}\equiv c/\Omega$ is the radius of the light cylinder. {Equation~(\ref{North1}) was first derived in \citet{Dimitropoulos2024} from Maxwell's equation $\nabla\times \vv{E} =-(1/c)\partial \vv{B}/\partial t$ and the condition that in steady-state in the pulsar co-rotating frame, $\partial \vv{B}/\partial t=-\nabla\times(r\sin\theta\ \Omega\ \hat{\phi}\times \vv{B})$.}
We emphasize that under steady-state ideal force-free conditions, electric fields have only a poloidal component along the $(r,\theta)$ meridional plane. Along the midplane of the ECS, $\vv{E} = \vv{B}=0$ because of the full field reversal that takes place across a Harris-type current sheet.
The drift velocity on both sides of the ECS is equal to
\begin{equation}
\vv{v}_{\rm ECS\ drift}=\frac{\vv{E}\times \vv{B}}{B^2}c=\frac{(-\vv{E})\times (-\vv{B})}{B^2}c.
\end{equation}
The result of both surfaces of the ECS drifting with $\vv{v}_{\rm ECS\ drift}$ is the corotation of the ECS pattern with the central star.

Next we consider a real non-ideal ECS with reconnection and energy dissipation taking place inside it. We also continue to assume that this ECS is embedded inside an ideal FFE magnetosphere where eq.~(\ref{North1}) is satisfied everywhere. The fact that there is reconnection across the ECS means that an additional small component $\vv{\delta B}$ of the magnetic field develops normal to the current sheet. This component of the  magnetic field crosses the ECS normally in the direction that is opposite to $\hat{n}$ since magnetic field lines reconnect from the north to the south. We notice that $\vv{\delta B}$ is absent in local PIC simulations of reconnection that start from a Harris-type current sheet. As we show below, this field component most probably stabilizes the ECS against the tearing instability.

We  estimate next the reconnecting magnetic field component by considering the rate of energy dissipation in a relativistic current sheet.  Numerical simulations tell us that energy is transferred to the ECS at a velocity of about $0.1c$ \citep[e.g.,][]{Hakobyan2023}.  That would imply that an extra electric field component $\vv{\delta E}$ develops above, inside, and below the ECS such that $\vv{\delta E}\times \vv{B}/B^2=f$ where $f\approx 0.1$.
Above and below the ECS steady-state FFE conditions apply, according to which the electric field can only have a poloidal component there. We therefore set
\begin{equation}
\vv{\delta E} \approx  f\ [\hat{n}\times \vv{B}-\hat{\phi}\   ([\hat{n}\times \vv{B}]\cdot\hat{\phi})].\ 
\label{North21}
\end{equation}
This generates an extra drift velocity of approach toward the ECS from both sides, namely
\begin{eqnarray}
\vv{v}_{\rm drift\ from\ north
}&=&\frac{\vv{\delta E}\times \vv{B}}{B^2}c\sim -fc,\ \hat{n}\nonumber\\
\vv{v}_{\rm drift\ from\ south}&=&\frac{\vv{\delta E}\times (-\vv{B})}{B^2}c\sim +fc\ \hat{n},
\end{eqnarray}
wherever $B_{\!p}\ll B_\phi$, and $f\approx 0.1$.
From eq.~(\ref{North1}) we also deduce that
\begin{eqnarray}
\vv{\delta B}_{p} &\equiv& \hat{\phi}\times \vv{\delta E}\left(\frac{R_{\rm lc}}{r\ \sin\theta}\right)\ \mbox{and that}\nonumber\\
\vv{\delta B} &\equiv& \vv{\delta B}_{p}-\frac{n_\phi}{n_{\!p}}\delta B_{p}\hat{\phi}
\label{North22}
\end{eqnarray}
so that $\vv{\delta B}$ is antiparallel to $\hat{n}$. Here $n_{\!p}
\equiv (n_r^2+n_\theta^2)^{1/2}$ and $\delta B_{\!p}\equiv (\delta B_{r}^2+\delta B_{\theta}^2)^{1/2}$. The negative sign is there because $\vv{\delta B} \cdot \hat{n}=-\delta B<0$, where $\delta B\equiv (\delta B_{r}^2+\delta B_{\theta}^2+\delta B_{\phi}^2)^{1/2}$. Therefore, along the midplane of the ECS two extra field components develop, namely $\vv{\delta B}$ and $\vv{\delta E}$ (see figure~\ref{ECS}). We have considered here the limit where $\vv{\delta B}$ is just a small perturbation to the overall ideal steady-state FFE structure of the pulsar magnetosphere ($\delta B\ll B$). Under this assumption, the overall geometry of the ECS remains unchanged, only it now contains an extra component of the magnetic field $\vv{\delta B}$ above, inside, and below the ECS that is normal to the ECS. It also contains an extra perpendicular poloidal component of the electric field $\vv{\delta E}$ above, inside, and below the ECS such that $\delta E\ll E$. Furthermore,
\begin{equation}
\frac{\delta E}{\delta B}=\left(\frac{r\ \sin\theta}{R_{\rm lc}}\right)n_{\!p}<\left(\frac{r\ \sin\theta}{R_{\rm lc}}\right)\ ;
\label{EB}
\end{equation}
therefore, particles along the midplace of the ECS travel in orthogonal electric and magnetic fields, and when ${(r\ \sin\theta / R_{\rm lc})n_{\!p}>1}$, the electric field becomes greater than the magnetic field and particles are accelerated along the direction of $\vv{\delta E}$. If the vector normal to the ECS has no $\phi$ component, this condition is satisfied right outside the light cylinder. In general, the ECS is expected to have a trailing spiral shape (i.e., $n_\phi >0$ and $n_{\!p}<1$); therefore, the electric field becomes greater than the magnetic field a small distance outside the light cylinder and beyond. 

Recent 3D PIC simulations of reconnection in pair plasmas  \citep[e.g.,][]{Zhang2021,Stathopoulos2024} have shown that there exist two distinct modes of acceleration of particles in a reconnecting current sheet. A fraction ($\sim 6\%$) of the pairs entering the current sheet meander (i.e. take a winding path)  between the two sides of the reconnection layer and undergo active acceleration every time they find themselves inside the current sheet. These are called free particles. The remaining  $94\%$ of the pairs are trapped in the current sheet and are accelerated in its midplane. We  now consider the two particle categories independently. {In the present work we do not discuss the origin of the particles that populate the ECS. They may originate in the central star, or in pair-formation gaps in the outer magnetosphere. Whatever their origin, they appear in numbers that are sufficient to support the magnetospheric electric charges and electric currents required by the ideal force-free conditions that we assume in the present work.}

\subsection{Trapped particles}

If the ECS does not develop tearing instability, charged particles in the ECS midplane will move under the combined effect of the perpendicular electric and magnetic fields, respectively $\vv{\delta E}$ and $\vv{\delta B}$. If $\delta E<\delta B$, they will follow a non-accelerating drift motion along $\vv{\delta E}\times \vv{\delta B}$. Wherever $\delta E>\delta B$, particles drift with velocity,
\begin{equation}
\vv{v}_{\rm ECS\ midplane\ drift}=\frac{\vv{\delta E}\times \vv{\delta B}}{\delta E^2}c,
\end{equation}
and are accelerated in the direction of $\vv{\delta E}$. Particle acceleration and dissipation will only take place outside the light cylinder. As we show in section~4 below, particles in the midplane of the ECS are quickly accelerated to relativistic velocities before they exit the ECS midplane. If positrons attain a component of the velocity equal to $\beta c$ along $\vv{\delta E}$, we assume that 
\begin{equation}
\left|\vv{v}_{\rm ECS\ midplane\ drift}\pm \beta c\ \vv{\delta E}/\delta E \right| \approx c ,
\label{drift1}
\end{equation}
from which we deduce that
\begin{equation}
\beta = \sqrt{1-\left(\frac{R_{\rm lc}}{r\ \sin\theta\ n_{\!p}}\right)^2}
\label{beta1}
\end{equation}
wherever $R_{\rm lc}/(r\ \sin\theta\ n_{\!p})\leq 1$. 
This constrains the radius where you can start to radiate beyond the light cylinder depending on the value of $n_{\!p}$.
The plus sign corresponds to positrons that are accelerated outward, and the minus sign to electrons that are accelerated inward. This determines the direction along which the accelerated particle emits radiation. In the radiation-reaction limit, the power of high-energy radiation per particle is equal to 
\begin{equation}
\mbox{Radiation power per particle}\propto e\beta c\ \delta E\ .
\label{power}
\end{equation}
The last thing that we need in order to generate high-energy sky-maps is the number of particles that radiate per unit surface of the ECS. An estimate of this is the surface number density $\sigma$ that corresponds to the surface charge density $\sigma_e$ of the ECS, namely
\begin{equation}
\sigma\approx \frac{\sigma_e}{e}= \frac{2 \vv{E}\cdot \hat{n}}{e}\ .
\label{sigma}
\end{equation}
This approximation assumes that there are not many more extra pairs in the ECS. {The true number density of radiating particles is likely to be higher since some multiplicity of pairs is required to support the global electric current and electric charge.
}

The above discussion is valid if the ECS does not develop tearing instability. We must understand that the $\vv{\delta E}$ component of the electric field that is responsible for bringing in Poynting flux $\vv{\delta E}\times\vv{B}$ from both sides of the current sheet toward its midplane originates in the ideal force-free magnetosphere outside the ECS. In a real pulsar magnetosphere, every magnetic field line originates on the surface of the central star, and therefore, as it is brought in by the $\vv{\delta E}$ component of the electric field that develops during reconnection, the magnetic field from the north magnetic hemisphere is connected to the magnetic field from the south magnetic hemisphere. This generates the extra consistent component $\vv{\delta B}$ of the magnetic field that is normal to the current sheet and points in the direction from the north to the south hemisphere. In general, such a normal component of the magnetic field stabilizes the current sheet against the tearing instability if
\begin{eqnarray}
\frac{\delta B}{B}&=&\frac{\delta E R_{\rm lc}}{r\ \sin\theta\ n_{\!p}B}\approx \frac{f R_{\rm lc}}{r\ \sin\theta\ n_{\!p}}\nonumber\\
&\sim& 0.1\gg S^{-3/4}\ \mbox{around the light cylinder}
\label{stabilization}
\end{eqnarray}
 \citep[e.g.,][]{CSstability}. Here, $S\equiv Lc/\eta_m$ is the Lundquist number in the ECS ($L$ is the half-thickness of the ECS, and $\eta_m$ the magnetic diffusivity). In astrophysical plasmas, $S\gg 10^{5}$ \citep[e.g.,][]{Zanotti2011}, and therefore we conclude that realistic ECSs are probably stable against the tearing instability, at least within a distance of a few times the size of the light cylinder. This is not the case, however, in current global numerical PIC simulations, where $\eta_m\approx \delta (0.1 c)$ is the numerical magnetic diffusivity with $\delta$ being the numerical grid resolution across the ECS, and $0.1 c$ being the approach velocity toward the ECS. Therefore, $S\approx 10 L/\delta$ is of order 10 in global PIC simulations. It is remarkable that in \cite{Soudais2024} the ECS is clearly stable inside about $1.2R_{\rm lc}$ where $\delta B/B$ is of order unity, and becomes unstable beyond that distance where $\delta B/B$ decreases significantly. We also note that local reconnection simulations that start from a Harris-type field configuration that does not contain a large-scale normal field component are always unstable to the tearing instability. We just argued that this is probably not the case in pulsar magnetospheres.

Nevertheless, if the ECS indeed manifests tearing instability,  magnetic X-points will develop in its interior (i.e., points where $\vv{B}=0$ and $\vv{\delta E}\neq 0$), and as long as particles stay near an X-point they will be accelerated in the direction of the electric field $\vv{\delta E}$.	This direction is different from the direction that we discussed above. We note however that, as $r\rightarrow\infty$, $\beta\rightarrow 1$ and, in both cases, particles move and radiate along $\vv{\delta E}$. We thus expect that the corresponding sky maps will be rather similar.

\subsection{Free particles}

This second distribution consists of particles that meander between the two sides of the reconnection layer and undergo active acceleration by the midplane electric field $\vv{\delta E}$ each time they cross the midplane of the current sheet. In that case, once they find themselves outside the ECS the relativistic particles attain a component of the velocity $\pm \beta' c$ along $\vv{B}$ such that
\begin{equation}
\left|\vv{v}_{\rm ECS\ drift}\pm\beta' c\ \vv{B}/B\right| \approx c 
\label{drift2}
\end{equation}
from which we can deduce that
\begin{equation}
\beta' = \sqrt{1-\left(\frac{r\ \sin\theta}{R_{\rm lc}}\right)^2\frac{1}{1+B_\phi^2/B_{\!p}^2}}\ .
\label{beta2}
\end{equation}
{It is interesting to note here the difference between eqs.~(\ref{beta1}) and (\ref{beta2}), and the similarity between eqs.~(\ref{drift1}) and (\ref{drift2}). As $r\rightarrow\infty$, $\beta\rightarrow 1$ and the positrons are accelerated outward along $\vv{\delta E}$, which becomes asymptotically radial, while $\beta'\rightarrow 0$ and particles drift outward with $\vv{v}_{\rm ECS\ drift}$, which also becomes asymptotically radial. Nevertheless, as we show below, the corresponding sky maps are rather different.}

\section{Sky maps}

We obtain high-energy sky maps as follows. We divide the ECS that we obtained in our simulation into small areas $\delta S$ around points $(r,\theta,\phi)$. Each such area contains a number of radiating particles equal to $\delta S \sigma$ (eq.~\ref{sigma}). Each particle radiates along a particular direction $\hat{\nu}=(\nu_r,\nu_\theta,\nu_\phi)$ that is determined by the specific particle acceleration and radiation model that we invoke, with an intensity that is proportional to the accelerating electric field component $\delta E$ (eq.~\ref{power}). We bin the radiation coming from each point $(r,\theta,\phi)$ of the ECS along $\theta$ and along a sky map phase equal to
\begin{equation}
-\phi-\delta\phi-r\Omega \nu_r/c\ ,
\label{phaseSM}
\end{equation}
where
\begin{equation}
\delta\phi\equiv {\rm arctan2}\left(\nu_\phi,\nu_r\sin\theta+\nu_\theta\cos\theta\right)\ .
\end{equation}
We note that the function ${\rm arctan2}$ yields an angle that lies between $0$ and $2\pi$ depending on the signs of $\nu_\phi$ and $\nu_r\sin\theta+\nu_\theta\cos\theta$. Here we explain how we calculate the phase of the radiation. We first consider the plane that contains the observers at various inclinations $\theta$ and the axis of rotation $\hat{z}\parallel\vv{\Omega}$. Phase zero of the sky map corresponds to the rotation phase when the magnetic axis $\vv{\mu}$ lies on that plane. This is when an observer at $\theta=\pi/2-\lambda$ observes the radio pulse coming from the north pole of the star, modulo the light travel time from the stellar surface to the observer that is the same in all our calculations. Obviously, the radiating point under consideration passes through the observer's plane at a phase $\phi$ before the radio pulse from the north pole. At that moment, the radiation direction $\hat{\nu}$ still does not lie on that plane. The star needs to turn an extra azimuthal angle $\delta\phi$ in order for the emitted radiation to be observed by one of our observers. We note that $\phi+\delta\phi$ is equal to the angle $\omega$ shown in figure~2 of \cite{Cerutti2016}. Radiation from that point at that phase of the rotation will reach the observer some time $\vv{r}\cdot\hat{\nu}/c$ {earlier} (or {later} if the latter dot product is negative) than a pulse emitted at the same time from the stellar surface at $r_*\ll R_{\rm lc}$. This time advance/delay corresponds to an extra phase advance/delay equal to $r\nu_r/R_{\rm lc}$. All terms in eq.~(\ref{phaseSM}) are in agreement with \cite{Cerutti2016}.

In figure~\ref{skymaps} we show the high-energy sky maps that correspond to the ECS obtained in the numerical solution of \cite{Dimitropoulos2026}. 
Since we are not running a particle simulation (PIC), we implement various reasonable prescriptions for the acceleration of particles.
The light curves that correspond to our sky maps are in general very sharp because our theoretical ECS has zero thickness.\footnote{{Figure~\ref{ECS} shows a closeup cross section of a current sheet with finite thickness $h$ which for the purpose of calculating radiation patterns is much smaller than the typical size of the ECS around the light cylinder.}}
In panel~(1) we assume that the positrons are trapped and accelerated in the midplane of the ECS wherever $\delta E>\delta B$ there, and radiate in the direction of $\vv{v}_{\rm ECS\ midplane\ drift}+ \beta c\ \vv{\delta E}/\delta E$ (eq.~\ref{drift1}). The electrons radiate in the direction with the negative sign. Their contribution may account for the Vela $\gamma$-ray interpulse emission between the two main positronic pulses \citep[e.g.,][]{Grenier1988}. The whole positron sky map is shifted in phase with respect to that expected from a split monopole emitting radially from the stellar surface (white dashed line; see next section) by roughly one-quarter of the period, but also the maximum latitudes reached by the radiation are significantly higher.
In panel~(2) we assume that a tearing instability develops in the ECS, X-points form, and the positrons that are trapped in the midplane of the ECS are accelerated in the direction of $\vv{\delta E}$, while the electrons are accelerated in the opposite direction. The sky maps are similar to the corresponding ones in the first panel, only they extend to even higher latitudes from the equatorial plane.
{The top two sky maps of figure~\ref{skymaps} were calculated inside distances of about $3R_{\rm lc}$ where $\delta E\gsim \delta B$. At larger distances $r\sin\theta\gg R_{\rm lc}$, the ECS becomes more and more spherical  in its largest part, and therefore $n_p\rightarrow 1$, $\beta\rightarrow 1$, $\delta E/\delta B\rightarrow \infty$, $\vv{\delta E}/\delta E\rightarrow \hat{r}$, and   the velocity and acceleration directions will become increasingly radial. 
}

\begin{figure}
		\centering
		\includegraphics[width=1.\columnwidth]{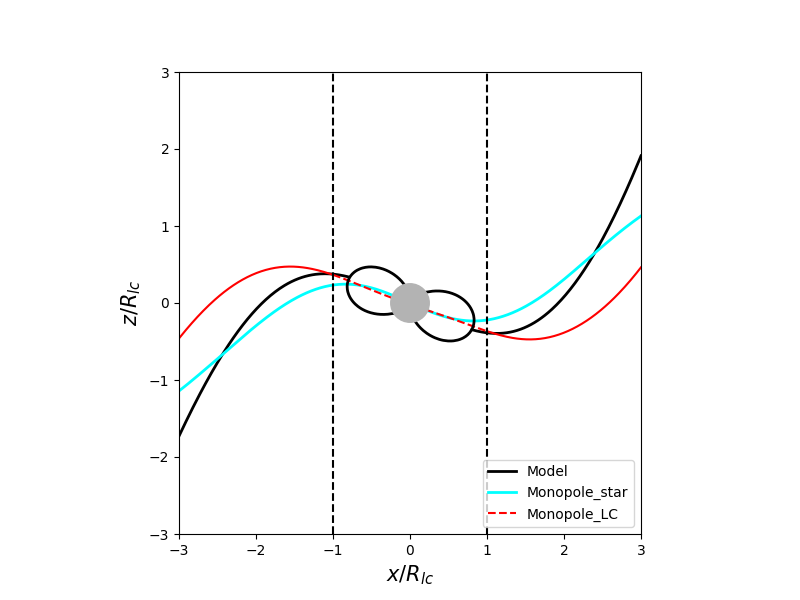}
		\caption{Cross section of the current sheet of our numerical solution (black line) along the plane containing the magnetic and rotation axes. {For comparison, the current sheet of the split monopole starting from the magnetic equator on the stellar surface (turquoise line) or from the cross section of the magnetic equator with the light cylinder (red line) are also shown. The dashed vertical lines indicate the light cylinder.} }
		\label{cross}
\end{figure}

Finally, in panel~(3) we assume that free particles follow the drift velocity of the magnetosphere and meander between the two sides of the ECS undergoing acceleration every time they find themselves inside it. Radiation is emitted along the direction of $\vv{v}_{\rm ECS\ drift}\pm\beta' c \vv{B}/B$ for both the positive and negative sign in eq.~(\ref{drift2}). We see that the sky map with the positive sign is much narrower than in the other panels,  due to the operation of the caustic effect (photons originating in different parts of the magnetosphere arriving in phase at the observer). This sky map is similar to the sky maps that the PIC simulations   generate, but is also {slightly} shifted in phase with respect to them. For example, the sky maps shown in \citep{Cerutti2025} have the same latitudinal extent and lie along the dashed white lines of the figure~\ref{skymaps} sky maps that correspond to a split monopole emitting from the stellar surface. In Figure~\ref{cross} we note the comparison between the ECSs in our numerical solution and in the split monopole {from the stellar surface and from the light cylinder}. Our sky map is shifted in phase between the white and red lines because of the extra phase shift in eq.~(\ref{phaseSM}) that is due to the time travel difference between photons emitted from the ECS beyond the light cylinder, and photons emitted from the stellar surface. 

{It is interesting that the sky maps that agree more closely  with those of the global PIC simulations are the ones that are generated with free particles. This would imply either that global PIC simulations miss the distribution of trapped particles because of limited spatial numerical  resolution across the ECS, either that the particles never become trapped along the ECS. As we show below, the latter is probably true and the analysis of \citet{Zhang2021,Stathopoulos2024} does not apply along the undulating (i.e., nonstationary) ECS of the pulsar magnetosphere.}

\section{The split-monopole approximation}

We believe that it is instructive to also obtain an approximate analytic form of the ECS according to the split monopole model and to generate high-energy sky-maps with which we will compare our results. We can directly check that the following expressions,
\begin{eqnarray}
B_r & = & B_*\frac{r_*^2}{r^2}\nonumber,\\
B_\phi & = & -B_*\frac{r_*^2}{R_{\rm lc}r}\sin \theta\nonumber,\\
E_\theta & = & B_\phi\ ,
\label{exact}
\end{eqnarray}
with $B_\theta=E_r=E_\phi=0$, 
consist an exact solution of the steady-state ideal force-free problem \citep{Michel1973}, 
\begin{equation}
\nabla\times\left[\vv{B}_{\!p}\left(1-\frac{r^2\sin^2\theta}{R_{\rm lc}^2}\right)+B_\phi \hat{\phi}\right]=\alpha \vv{B}\ \ \ \mbox{and}\ \ \ \vv{B}\cdot \nabla\alpha=0\ ,
\end{equation}
\citep[e.g.,][]{Endean1974,Mestel1975,Dimitropoulos2024}, with monopole boundary conditions on the surface of the star at $r=r_*$, namely
\begin{equation}
B_r(r_*) = B_*\ ,
\end{equation}
and $\alpha = -2\cos\theta/R_{\rm lc}$. 
We note that the above monopole solution crosses  the light cylinder smoothly. \cite{Bogovalov1999} showed that the split monopole where the radial magnetic field direction changes sign in the southern stellar magnetic hemisphere is equivalent to the above configuration only with the inclusion of a corotating undulating ECS separating the two magnetospheric hemispheres of opposite polarity. The shape of the split monopole ECS is given parametrically by 
\begin{equation}
\frac{{\rm d}r}{B_r}=\frac{r\ {\rm d}\theta}{B_\theta}
=\frac{r\sin\theta\ {\rm d}\phi}{B_\phi}
\end{equation}
for all magnetic lines of the monopole solution that originate on the stellar magnetic equator. According to eqs.~(\ref{exact}) above, $B_\theta=0$, and thus also ${\rm d}\theta=0$ and ${\rm d}r=R_{\rm lc}{\rm d}\phi$. The latter implies that  ${\rm d}\phi=2\pi\Rightarrow {\rm d} r=cP$, where $P=2\pi/\Omega$ is the pulsar period. In other words, the ECS is contained within latitudes $\pm \lambda$ from the rotational equator ($\lambda$ is the pulsar inclination angle), and along any such latitude, the distance between two successive crossings of the ECS is equal to $cP$. Equivalently, the ECS of the split monopole may be traced by shooting photons radially from the magnetic equator. \cite{Bogovalov1999} derived the equation that yields the position of the ECS of the split monopole, namely
\[
\eta(r,\theta,\phi;t) =
\]
\[
\sin\lambda \sin\left(\theta-\int_{r_*}^r\frac{B_\theta{\rm d}r}{r B_r}\right) \cos\left(\phi-\int_{r_*}^r\frac{B_\phi{\rm d}r}{r\sin\theta B_r}-\Omega t\right)+\cos\theta \cos\lambda
\]
\begin{eqnarray}
&=& \sin\lambda \sin\theta \cos\phi'+\cos\lambda \cos\theta 
\nonumber\\
&=& 0\ ,
\label{eta}
\end{eqnarray}
where the retarded azimuthal angle $\phi'$ is equal to
\begin{equation}
\phi'\equiv \phi+\frac{r-r_*}{R_{\rm lc}}-\Omega t
\end{equation}
We do not understand why \cite{Bogovalov1999} also included  a $B_\theta$ term in eq.~(\ref{eta}) and in his Fig.~4 since $B_\theta=0$ by definition in the split monopole. We  also note that we have cosine instead of sine in the $\phi$ term because of our different choice of the $\phi=0$ plane (in our case, at $t=0$ the $\phi=0$ plane contains the inclined magnetic axis). We can then directly calculate the normal unit vector to the ECS, namely
\begin{eqnarray}
\hat{n} &\equiv & (n_r, n_\theta, n_\phi)\nonumber\\
& \propto & -\nabla\eta\nonumber\\
& = &\left(\frac{\sin\lambda\sin\theta \sin\phi'}{R_{\rm lc}}\right.\ ,\nonumber\\
&& \left. \frac{-\sin\lambda \cos\theta \cos\phi'+\cos\lambda\sin\theta}{r}\ ,\frac{\sin\lambda\sin\phi'}{r}\right)\ .
\label{hatn}
\end{eqnarray}
The minus sign in eq.~(\ref{hatn}) was chosen such that the normal vector $\hat{n}$ points from the southern to the northern magnetic hemisphere as in the cartoon in Figure~\ref{ECS}.

\begin{figure*}
\centering
\begin{tabular}{c}
  \includegraphics[width=10.5cm,height=4.2cm,angle=0.0]{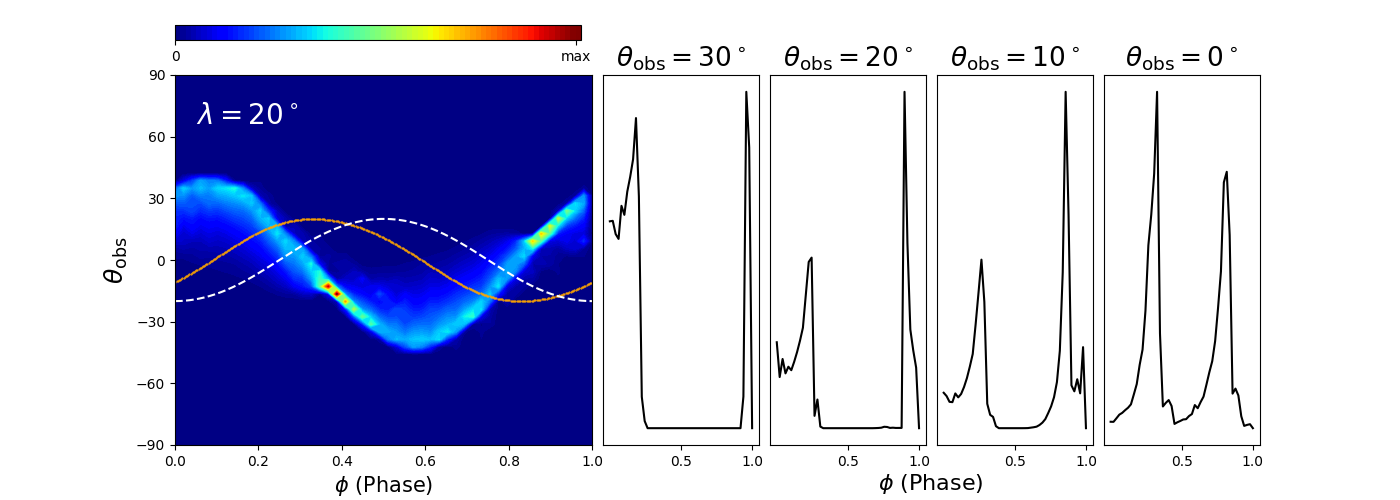}\hspace{-1cm}
  \includegraphics[width=10.5cm,height=4.2cm,angle=0.0]{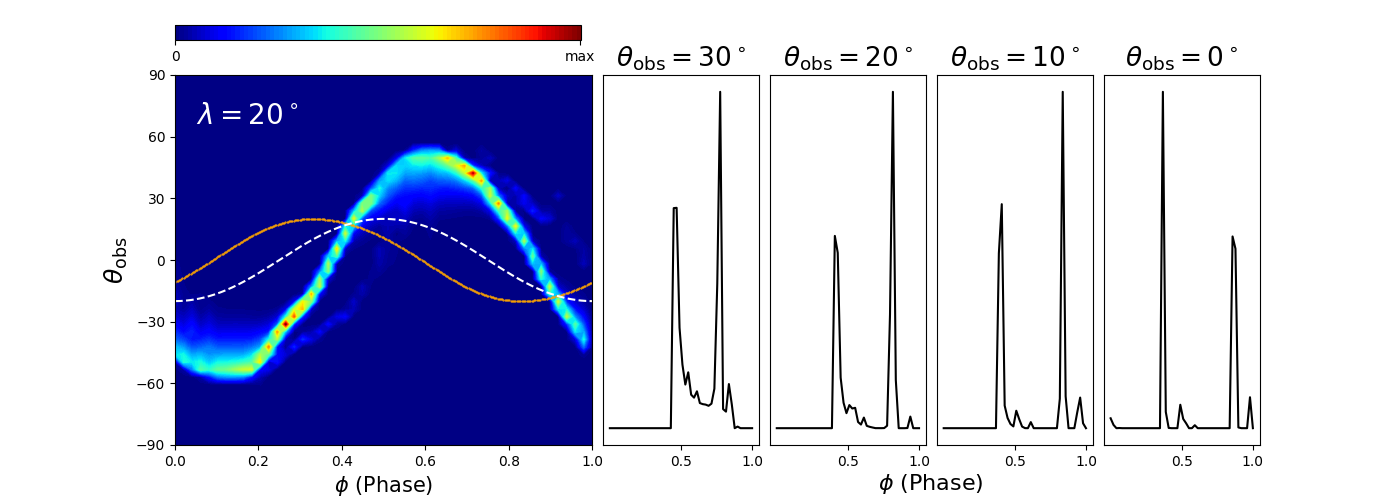} \\
\multicolumn{1}{c}{(1) Trapped particles in the ECS midplane radiating along $\vv{v}_{\rm ECS\ midplane\ drift}\pm\beta c\vv{\delta E}/\delta E$ for positrons/electrons respectively}\\
 \includegraphics[width=10.5cm,height=4.2cm,angle=0.0]{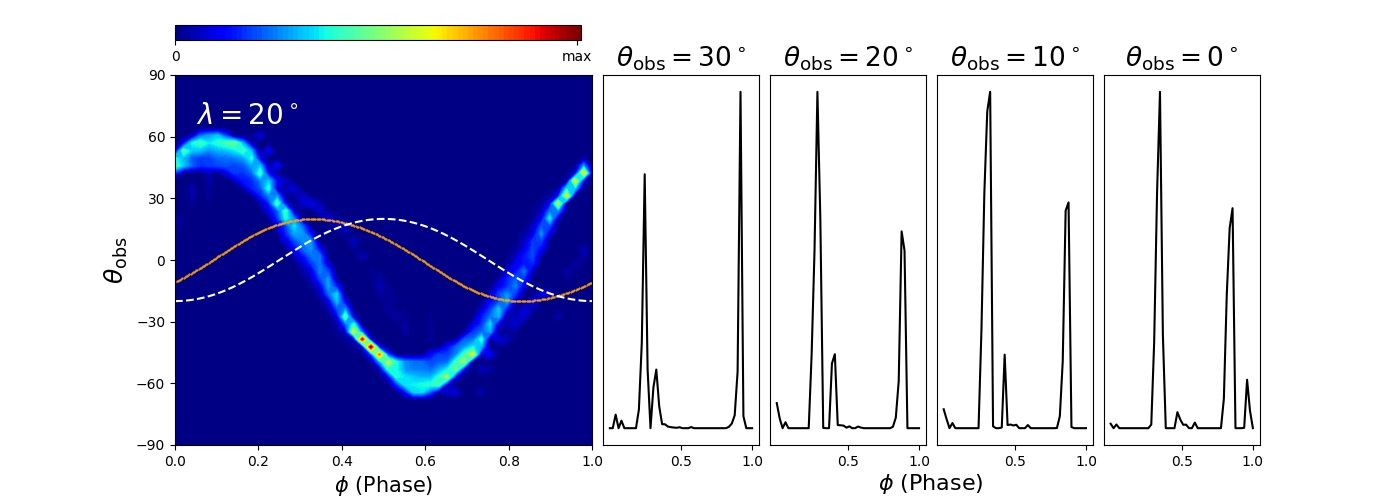}\hspace{-1cm} \includegraphics[width=10.5cm,height=4.2cm,angle=0.0]{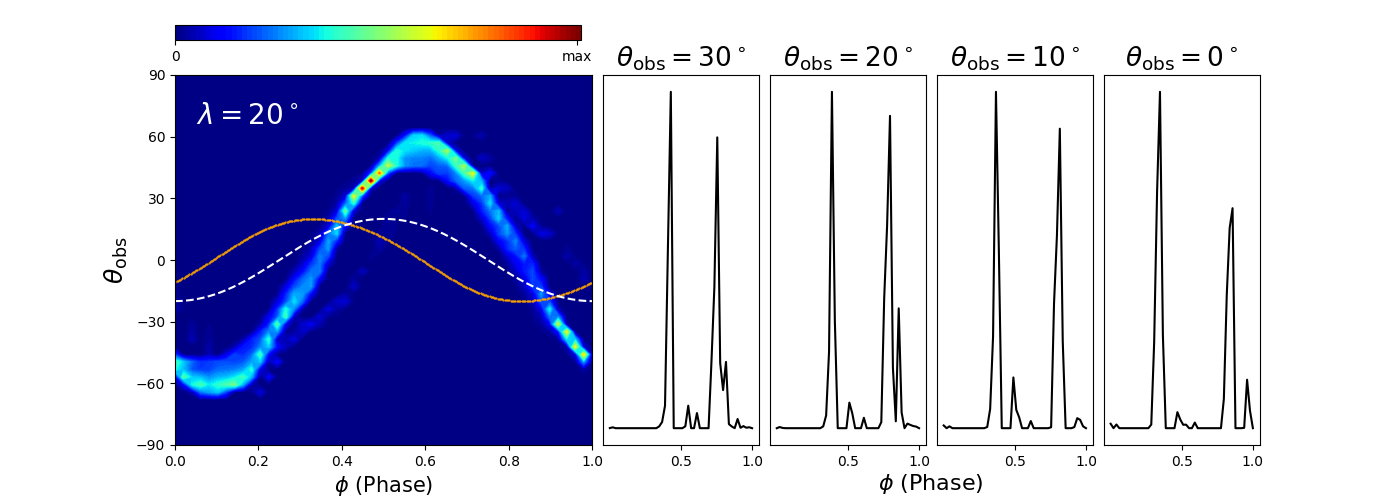} \\
\multicolumn{1}{c}{(2) Trapped particles in the ECS midplane radiating  along $\pm\vv{\delta E}$ for positrons/electrons respectively}\\
 \includegraphics[width=10.5cm,height=4.2cm,angle=0.0]{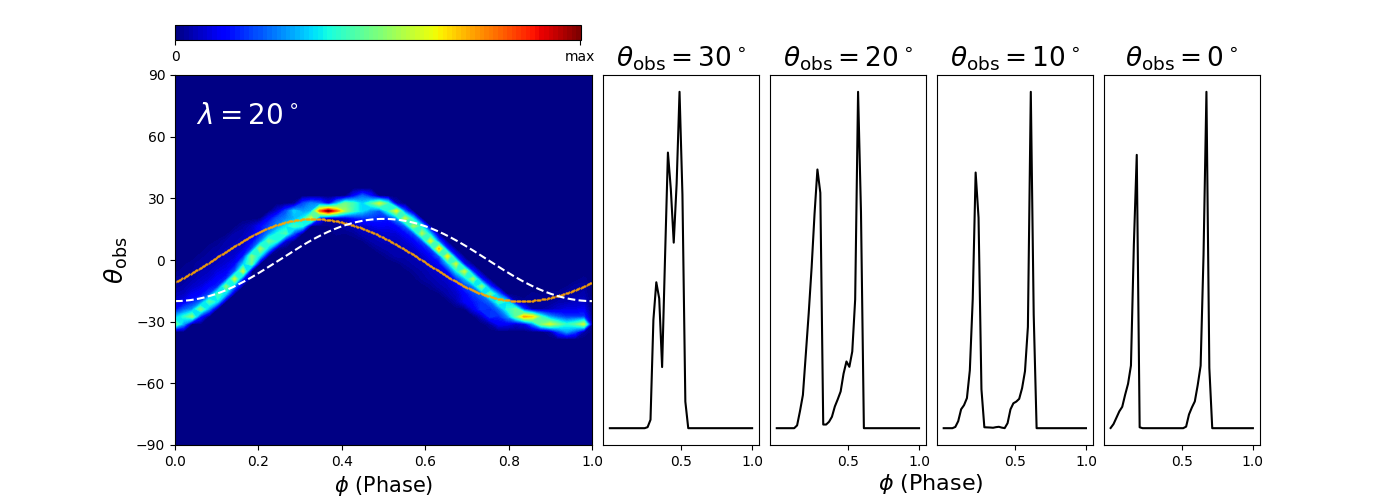} \hspace{-1cm}
 \includegraphics[width=10.5cm,height=4.2cm,angle=0.0]{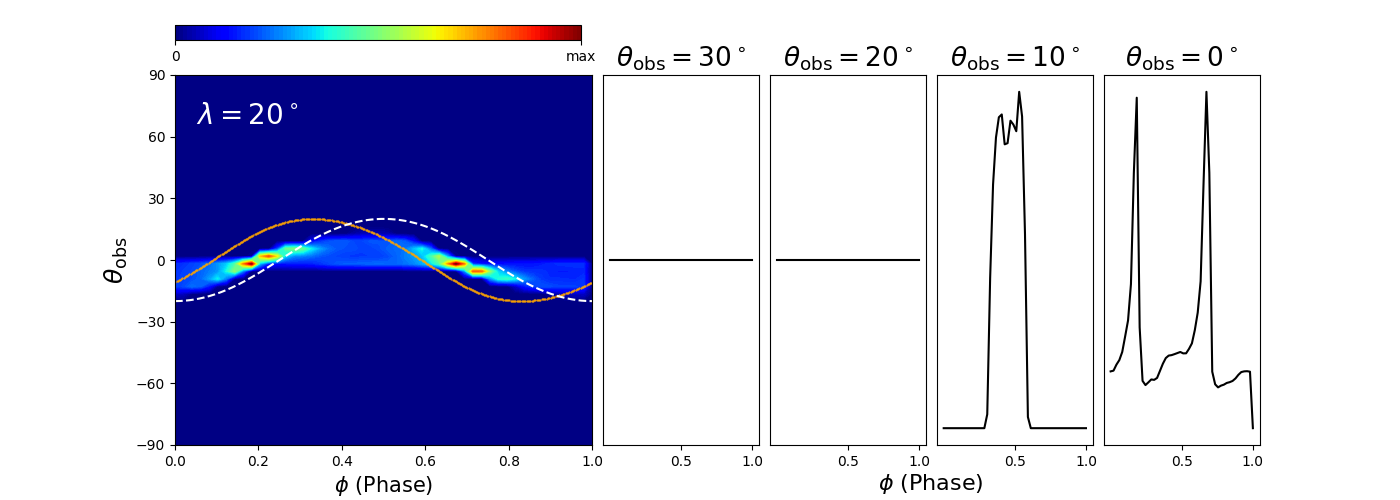} \\
\multicolumn{1}{c}{(3) Free particles radiating  along $\vv{v}_{\rm ECS\ drift}\pm\beta' c\vv{B}/B$ 
}
\end{tabular}
\caption{High-energy sky maps and sample light curves for pulsar inclination $\lambda=20^\circ$. {$\theta=0$ corresponds to an observer along the spin equator.} Phase zero corresponds to the arrival of the radio pulse emitted from the stellar magnetic poles. The white dashed line is the radial radiation from the ECS of a central split monopole from the stellar surface and beyond. The red line is the radial radiation from the ECS of a central split monopole from the light cylinder and beyond. We observe that in all cases except for (3), the radiation extends to latitudes higher than $\pm\lambda$. We also expect that since the ECS consists mostly of positrons, the right  plots in the top two panels that correspond to the electrons are expected to be much weaker (but not negligible) than the left plots, which correspond to the positrons.
}
\label{skymaps}
\end{figure*}

\begin{figure*}
\centering
\begin{tabular}{c}
  \includegraphics[width=10.5cm,height=4.2cm,angle=0.0]{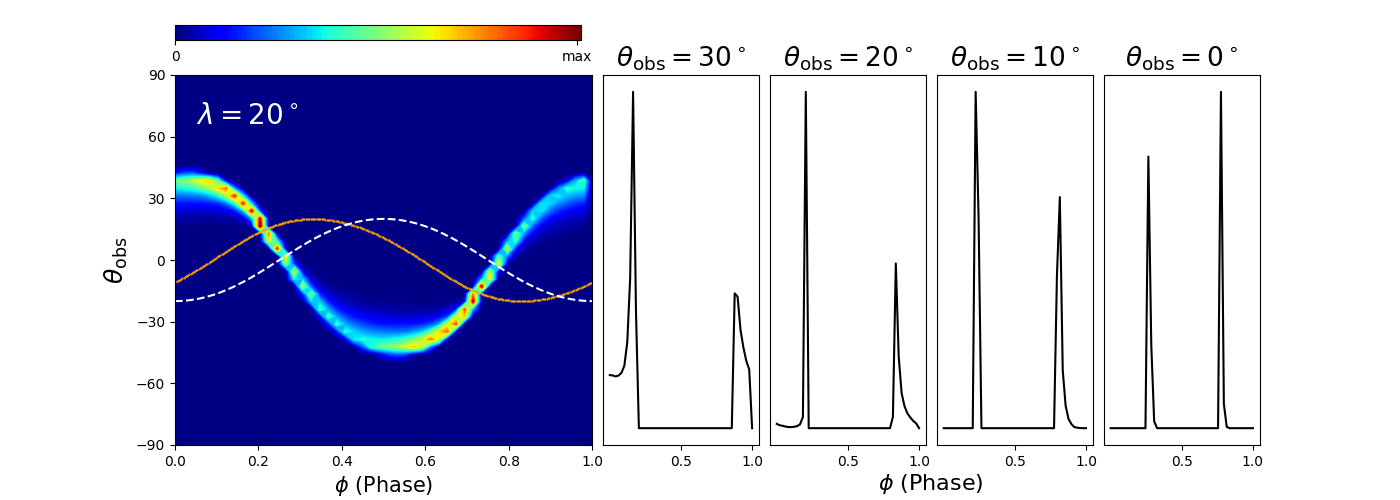}\hspace{-1cm}
  \includegraphics[width=10.5cm,height=4.2cm,angle=0.0]{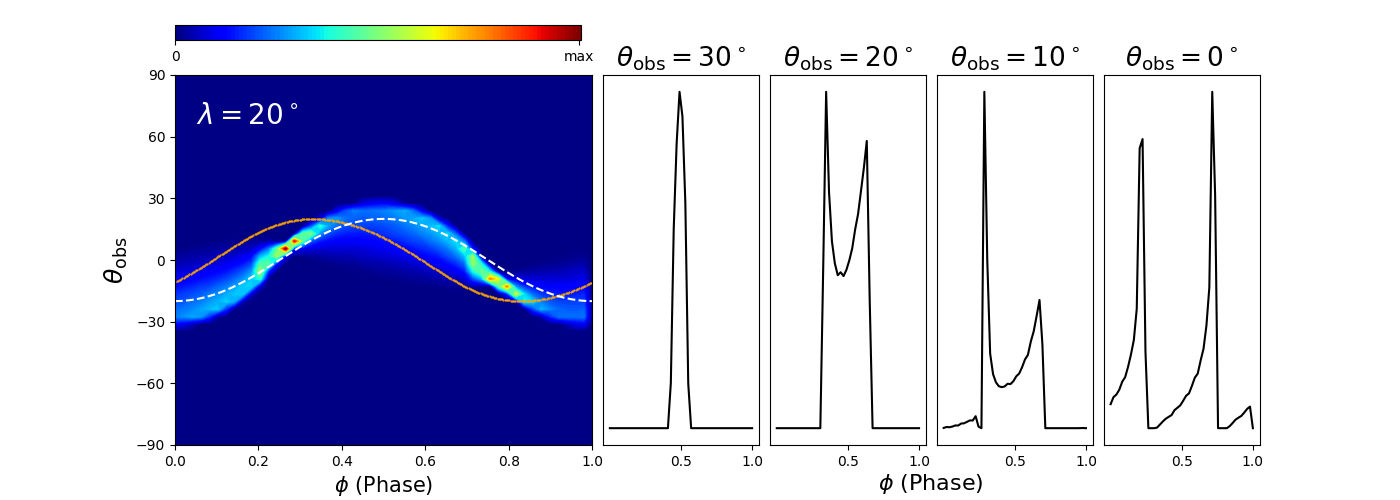} \\
\multicolumn{1}{c}{(1) Trapped particles in the ECS midplane radiating along $\vv{v}_{\rm ECS\ midplane\ drift}\pm\beta c\vv{\delta E}/\delta E$ for positrons/electrons respectively}\\
 \includegraphics[width=10.5cm,height=4.2cm,angle=0.0]{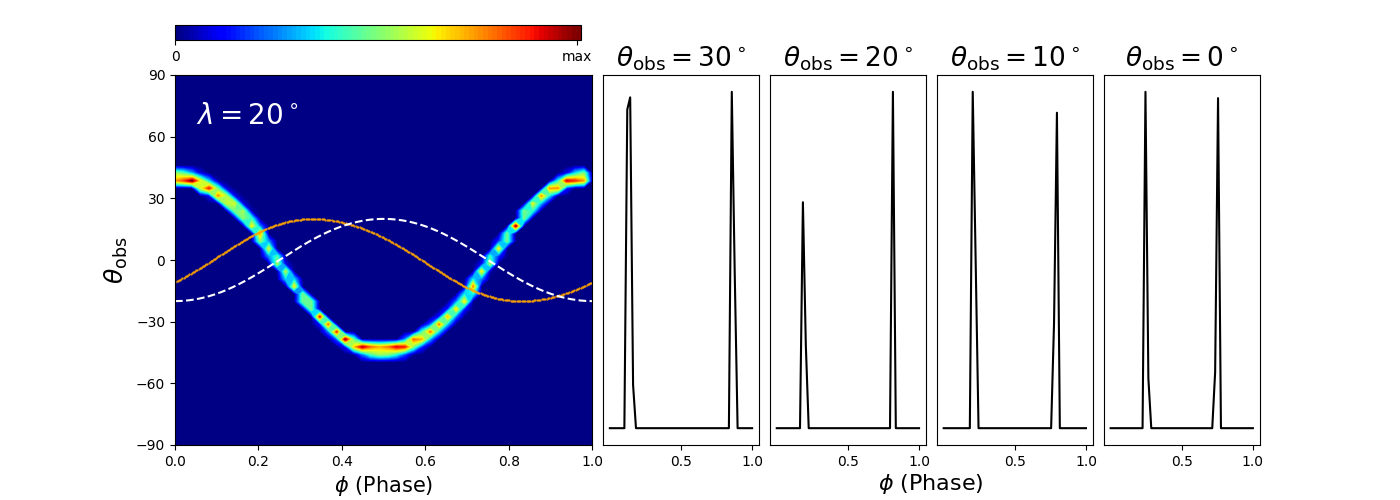}\hspace{-1cm} \includegraphics[width=10.5cm,height=4.2cm,angle=0.0]{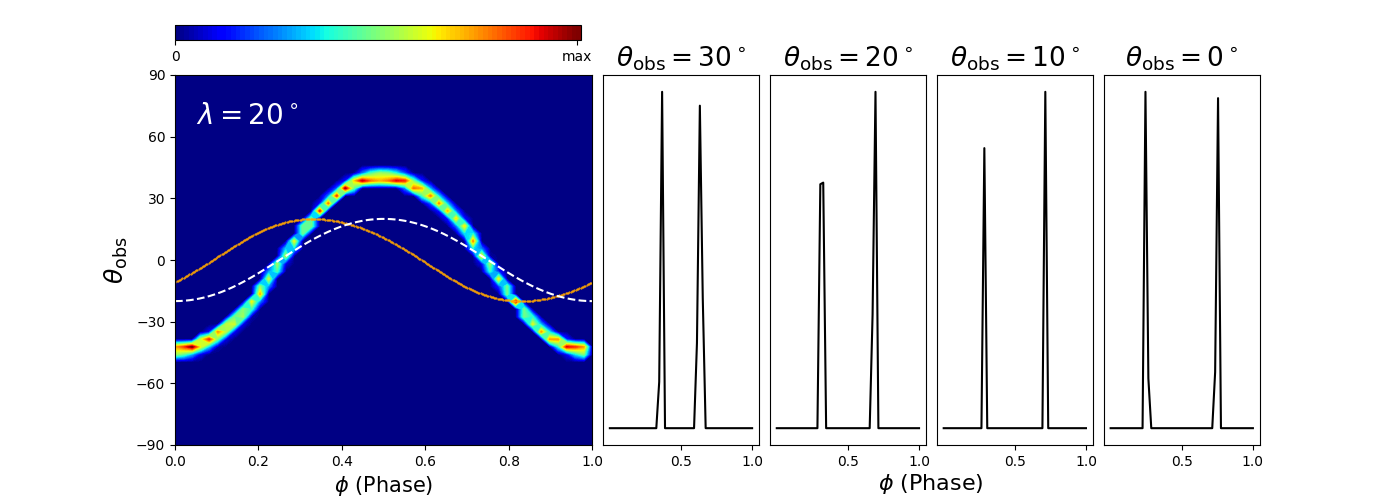} \\
\multicolumn{1}{c}{(2) Trapped particles in the ECS midplane radiating  along $\pm\vv{\delta E}$ for positrons/electrons respectively}\\
 \includegraphics[width=10.5cm,height=4.2cm,angle=0.0]{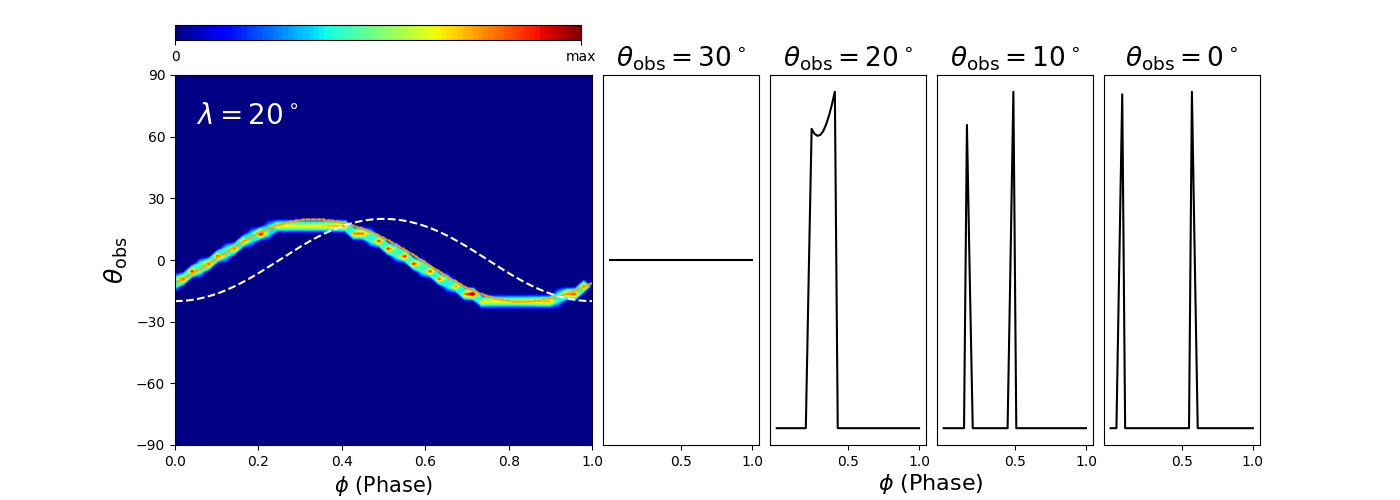} \hspace{-1cm}
 \includegraphics[width=10.5cm,height=4.2cm,angle=0.0]{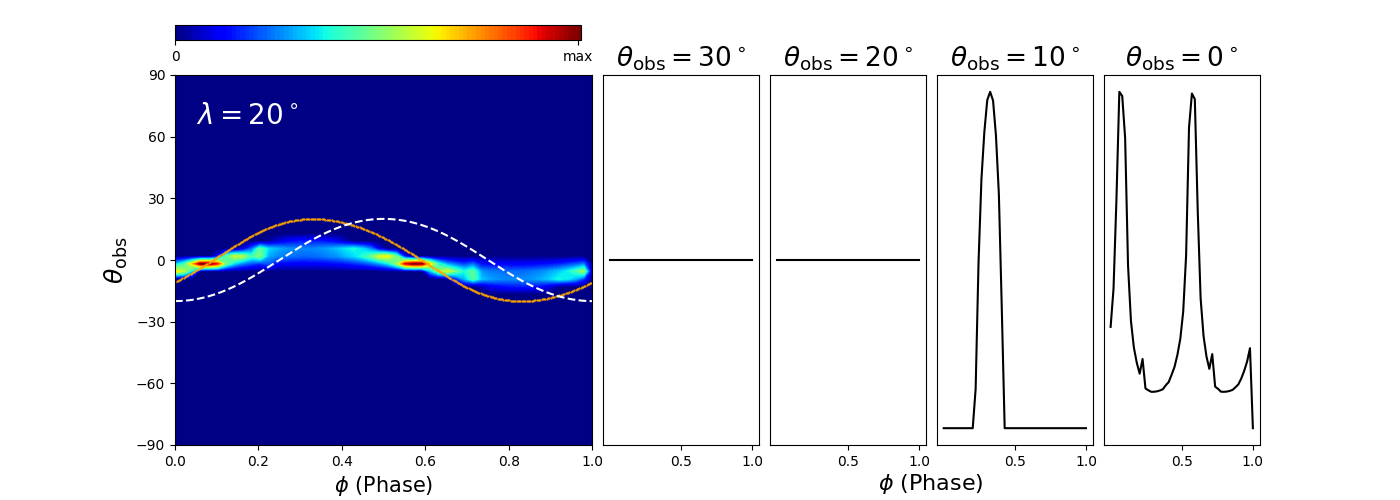} \\
\multicolumn{1}{c}{(3) Free particles radiating along $\vv{v}_{\rm ECS\ drift}\pm\beta' c\vv{B}/B$ beyond the light cylinder 
}\\
 \includegraphics[width=10.5cm,height=4.2cm,angle=0.0]{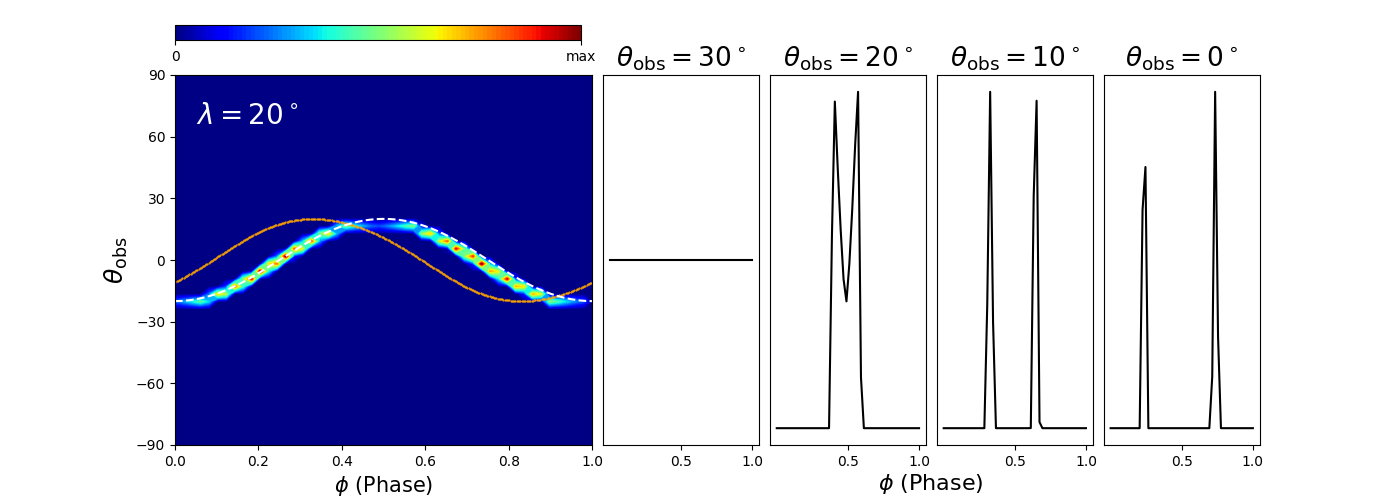} \hspace{-1cm}
 \includegraphics[width=10.5cm,height=4.2cm,angle=0.0]{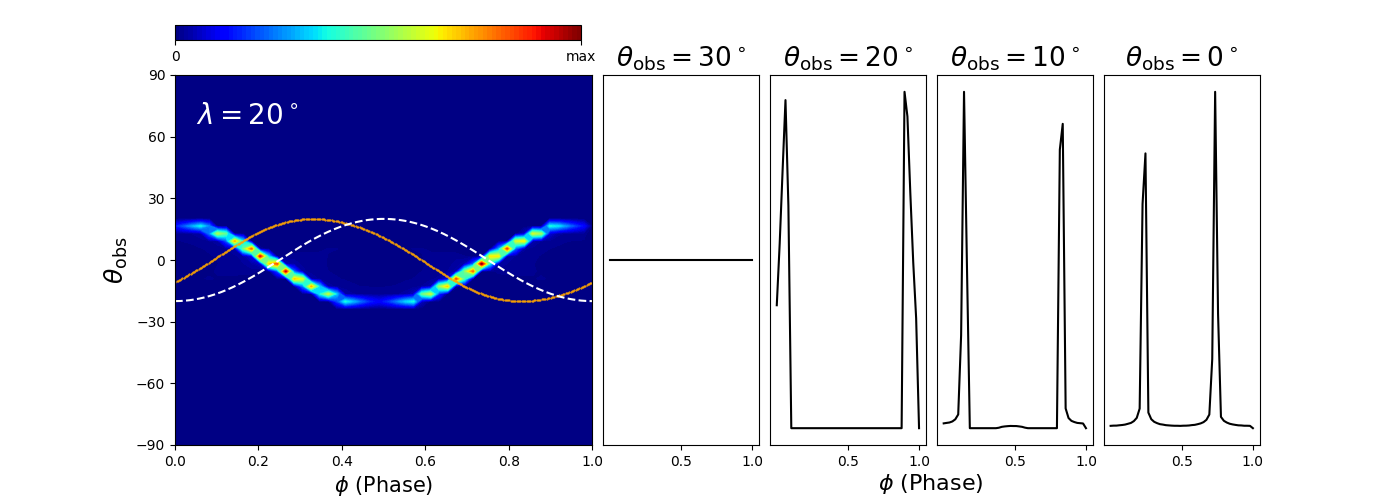} \\
\multicolumn{1}{c}{(4) Same as (3), only free particles start emitting from the stellar surface
}
\end{tabular}
\caption{Similar to Figure~\ref{skymaps} for the split monopole solution of section 3 at inclination $\lambda=20^\circ$. The light curves are very similar to those produced with our numerical solution. They are also narrower (more caustic-like). We see the clear shift from the red to the white  line when free particles in the ECS start emitting from the stellar surface.
}
\label{skymaps2}
\end{figure*}

We can now generate sky maps from the ECS of the split monopole (see figure~\ref{skymaps2}). If the ECS radiates radial photons from the stellar surface and beyond, they will be collected along the dashed white lines, which are obtained without intensity information. If the ECS radiates radial photons from the light cylinder and beyond, they will be collected along  the red lines (again without intensity information), which are not only shifted to the left, but also differ in shape from the white dashed lines. For radiation directions different from radial, the generation of sky maps is more complex, as described in the previous section. In general, the light curves are similar to those produced with our numerical solution in Figure~\ref{skymaps}. They are also narrower (more caustic-like).

\begin{figure*}
\centering
\begin{tabular}{c}
\hspace{-1cm}
 \includegraphics[width=12.5cm,height=4.2cm,angle=0.0]{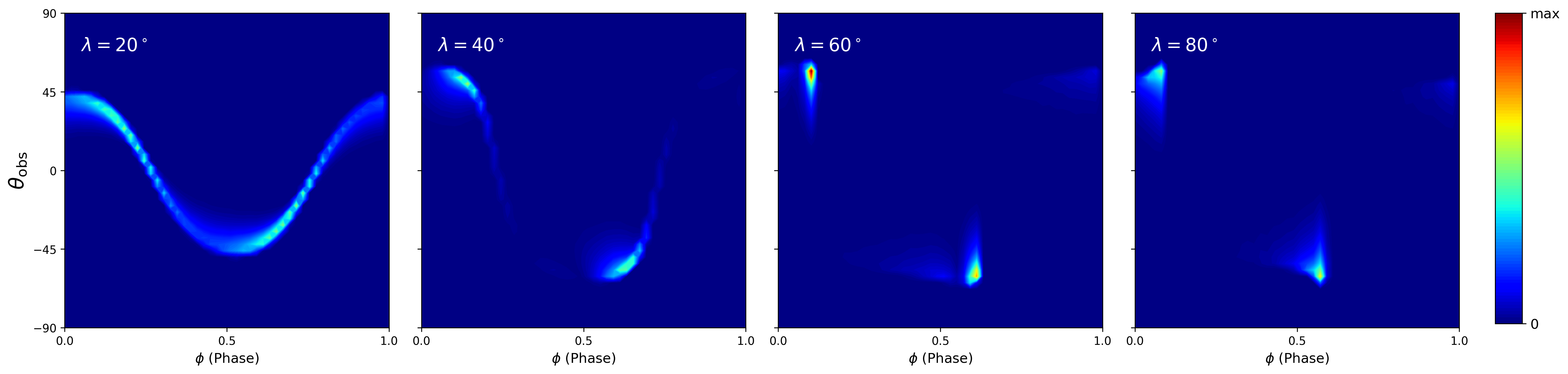}\hspace{-1cm}
 \includegraphics[width=8cm,height=4.2cm,angle=0.0]{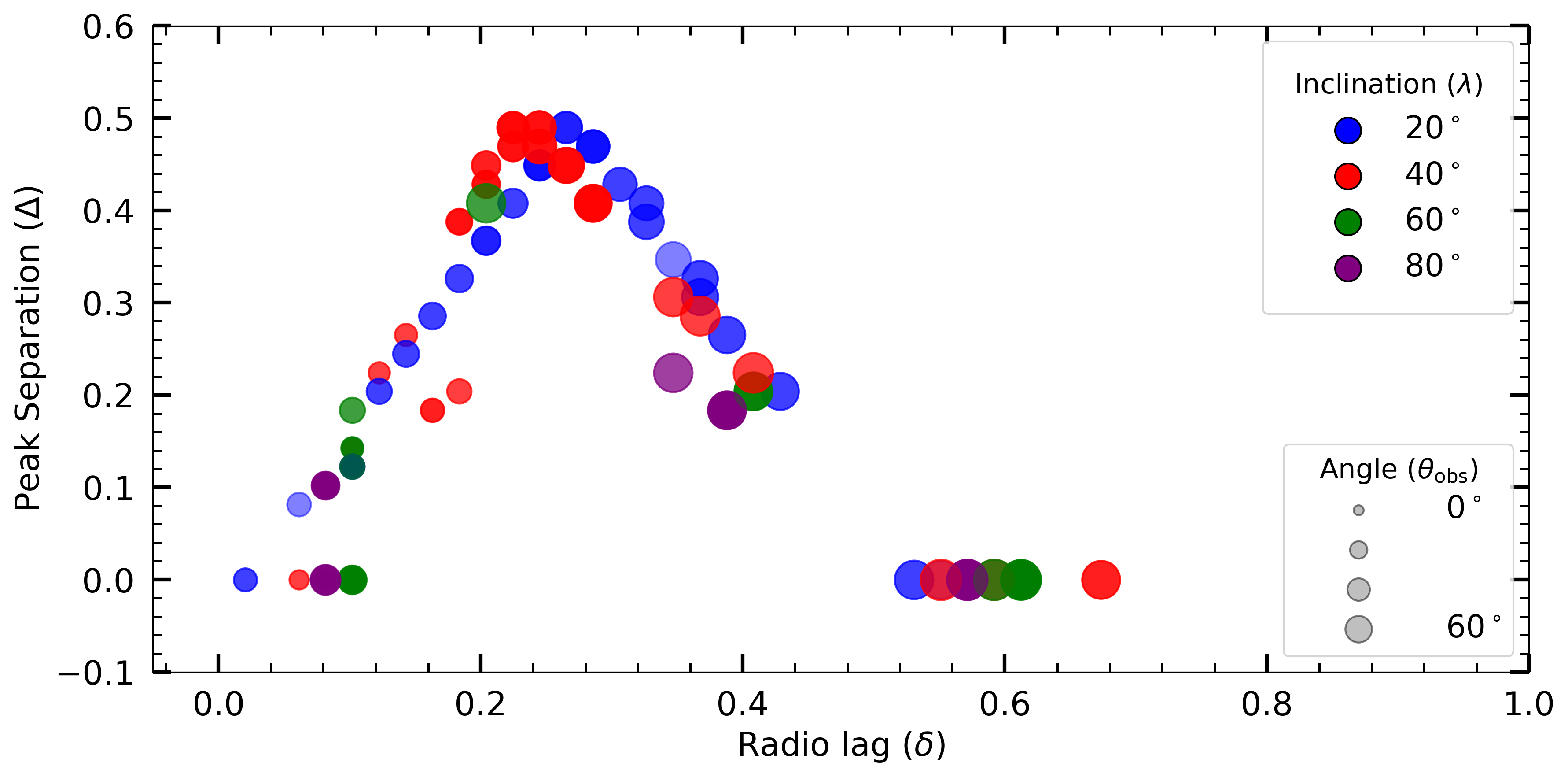} \\
\multicolumn{1}{c}{(1) Trapped positrons radiating beyond the light cylinder in the ECS of a split monopole.}\\
\hspace{-1cm}
 \includegraphics[width=12.5cm,height=4.2cm,angle=0.0]{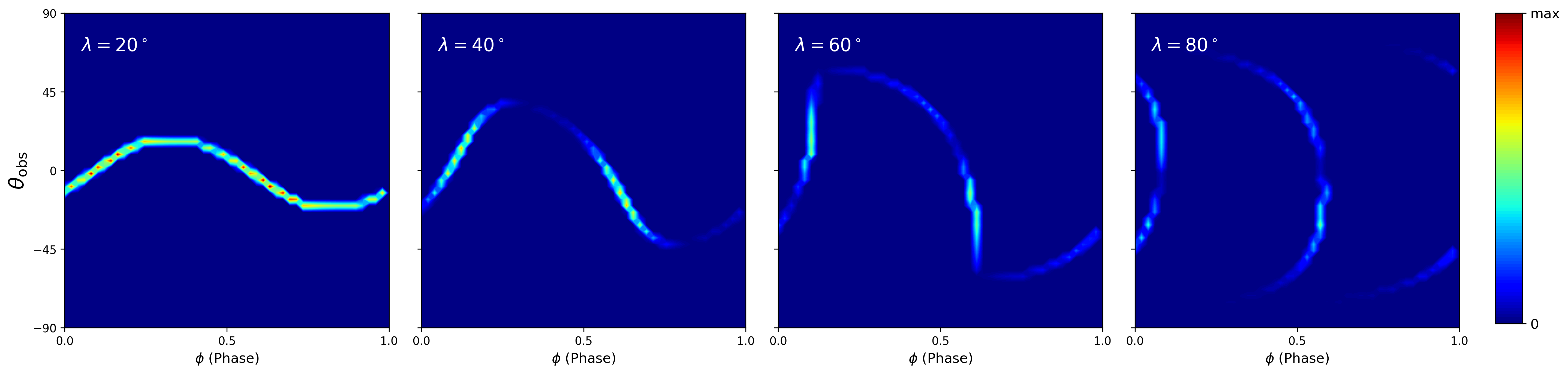}\hspace{-1cm}
 \includegraphics[width=8cm,height=4.2cm,angle=0.0]{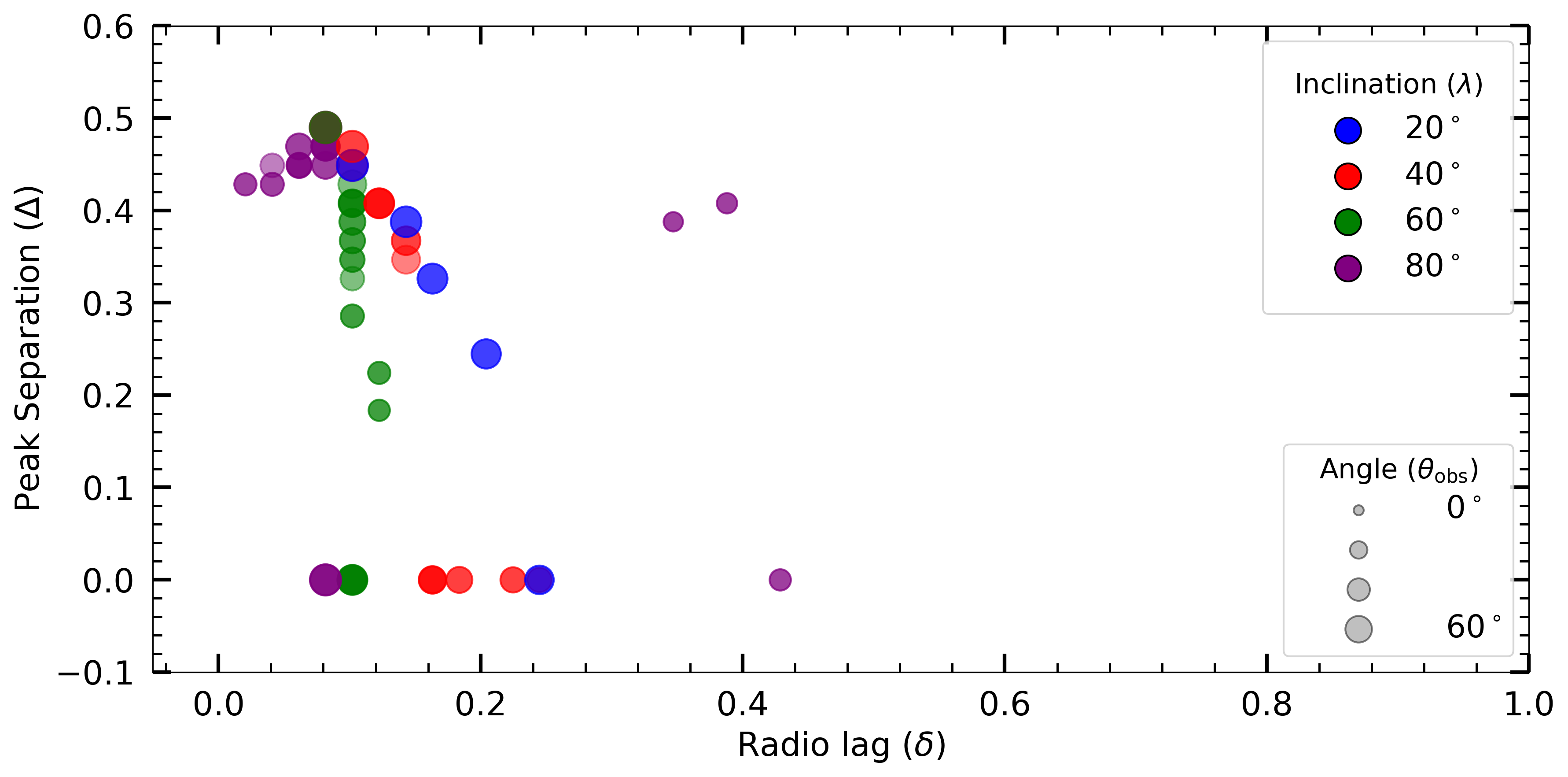}\\
\multicolumn{1}{c}{(2) Free positrons radiating beyond the light cylinder in the ECS of a split monopole.}
\end{tabular}
\caption{{Sky maps for the split monopole solution of section 3 at various pulsar inclinations $\lambda$ (left) and their corresponding $\Delta$~vs.~$\delta$ diagrams (right) for trapped (top) and free (bottom) positrons. The various inclination and observation angles $\lambda$ and $\theta_{\rm obs}$ are shown in the legends. The lower diagram matches more closely figs.~13 and 10 of  \citet{Kalapotharakos2023} and \citet{Smith2023}, respectively.}
}
\label{skymaps3}
\end{figure*}

One way to test our sky map models is to generate from them average diagrams of phase separation between main gamma-ray peaks $\Delta$ versus the phase lag of the first gamma-ray peak relative to the radio emission $\delta$, as  in \cite{Kalapotharakos2023}, among others, and compare them with the Third Fermi Pulsar Catalogue (3PC). We acknowledge that our global force-free model was only computed for one pulsar inclination angle and that we are not yet confident to present our results for higher angles. We thus decided to include computations for a few larger inclination angles using the split monopole ECS model. In figure~\ref{skymaps3} we obtain sky maps for the split monopole solution of section 3 at various pulsar inclinations $\lambda$ and their corresponding $\Delta$~versus~$\delta$ diagrams for positrons in the ECS radiating along $\vv{v}_{\rm ECS\ midplane\ drift}\pm\beta c\vv{\delta E}/\delta E$ (trapped) and along $\vv{v}_{\rm ECS\ drift}\pm\beta' c\vv{B}/B$ (free) beyond the light cylinder. The $\Delta$~versus~$\delta$ diagram, which matches more closely figs.~13 and 10  in respectively  \citet{Kalapotharakos2023} and \citet{Smith2023},  is the bottom one that corresponds to free positrons. This result confirms that  particles indeed do not become trapped along the ECS, in agreement with global PIC simulations.

The agreement of our $\Delta$~versus~$\delta$ diagram with 3PC is interesting. Previous model light curves using force-free magnetospheres produced radio lags that were larger than observed and this was used as a justification for dissipation in  global models that produced better agreement with observations \citep[e.g.,][]{Kalapotharakos2014,Cao2019,Cao2022,Cao2024}. Our simple split-monopole model with dissipation in the ECS produces an equally good fit with globally dissipative models.

\section{Physical parameters in the ECS}

We  now check whether  particles indeed reach the radiation-reaction limit inside the midplane of the ECS before the latter moves away and the particle is found outside the midplane.
We start with the basic quantitites of our physical system:
\[
R_{\rm lc}=\frac{cP}{2\pi}=4.78\times 10^8{\rm cm}\ \left(\frac{P}{0.1{\rm s}}\right),
\]
\[
B_{\rm lc}\sim B_* (r_*/R_{\rm lc})^3 \sim 1.59\times 10^4\ {\rm G}\ \left(\frac{B_*}{10^{12}\ {\rm G}}\right)\left(\frac{P}{0.1{\rm s}}\right)^{-3},
\]
\begin{eqnarray}
\delta B_{\rm lc}&=&\delta E_{\rm lc}\sim f B_{\rm lc}\nonumber\\
&\sim& 1.59\times 10^3\ {\rm G}\ \left(\frac{f}{0.1}\right)\left(\frac{B_*}{10^{12}\ {\rm G}}\right)\left(\frac{P}{0.1{\rm s}}\right)^{-3}\nonumber.
\end{eqnarray}
Here $r_*=1.2$~km. In the curvature radiation-reaction limit where ${ce\delta E\sim (2/3)e^2 c\gamma_{\rm curv}^4/R_{\rm lc}^2}$, the limiting Lorentz factor $\gamma_{\rm curv}$ is equal to
\begin{eqnarray}
\gamma_{\rm curv}&\sim& \left(\frac{3\delta E R_{\rm lc}^2}{2e}\right)^{1/4}\nonumber\\
&\sim& 3.26\times 10^{7}\ \left(\frac{f}{0.1}\right)^{1/4}\left(\frac{B_*}{10^{12}\ {\rm G}}\right)^{1/4}\left(\frac{P}{0.1{\rm s}}\right)^{-1/4}\ ,
\label{gammac}
\end{eqnarray}
the characteristic radiation energy is equal to
\begin{eqnarray}
\epsilon_{\rm curv}&=& \frac{3hc}{4\pi R_{\rm lc}}\gamma_{\rm curv}^3\nonumber\\
&\sim & 2.16\ {\rm GeV}\ \left(\frac{f}{0.1}\right)^{3/4}\left(\frac{B_*}{10^{12}\ {\rm G}}\right)^{3/4}\left(\frac{P}{0.1{\rm s}}\right)^{-7/4}
\label{radgammac}
\end{eqnarray}
($h$ is the Planck constant), while the acceleration length $l_{\rm acc}$ is equal to
\begin{eqnarray}
l_{\rm acc}&\sim& \frac{\gamma_{\rm curv} m_e c^2}{e\delta E}\nonumber\\
&\sim&
3.51\times 10^8{\rm cm}
\ \left(\frac{f}{0.1}\right)^{-3/4}\left(\frac{B_*}{10^{12}\ {\rm G}}\right)^{-3/4}\left(\frac{P}{0.1{\rm s}}\right)^{11/4}\nonumber\\
&\sim& 0.73\ R_{\rm lc}\ \left(\frac{f}{0.1}\right)^{-3/4}\left(\frac{B_*}{10^{12}\ {\rm G}}\right)^{-3/4}\left(\frac{P}{0.1{\rm s}}\right)^{7/4}\nonumber.
\end{eqnarray}
This acceleration length may be a significant fraction of the light cylinder radius, and thus, {as we argue above, the particles are probably not trapped in the midplane of the undulating ECS during their acceleration by the reconnection electric field $\vv{\delta E}$.}

Similarly, in the synchrotron radiation-reaction limit where $ce\delta E\sim (2/3)e^4\gamma_{\rm sync}^2 \delta B^2\sin^2\alpha/(m_e^2c^3)$, the limiting Lorentz factor $\gamma_{\rm sync}$ for pitch angles $\alpha\approx 90^\circ$ is equal to
\begin{eqnarray}
\gamma_{\rm sync}&\sim& \left(\frac{3m_e^2 c^4 \delta E}{2e^3 \delta B^2}\right)^{1/2}\nonumber\\
&\sim& 2.39\times 10^{6}\ \left(\frac{f}{0.1}\right)^{-1/2}\left(\frac{B_*}{10^{12}\ {\rm G}}\right)^{-1/2}\left(\frac{P}{0.1{\rm s}}\right)^{3/2}\ ,
\label{gammas}
\end{eqnarray}
the characteristic radiation energy is equal to
\begin{eqnarray}
\epsilon_{\rm sync} &=&\frac{3h}{4\pi}\gamma_{\rm sync}^2\frac{e\delta B}{m_e c}\nonumber\\
&\sim& 158\ {\rm MeV}\ ,
\label{radgammas}
\end{eqnarray}
while the acceleration length $l_{\rm acc}$ is equal to
\begin{eqnarray}
l_{\rm acc}&\sim& \frac{\gamma_{\rm sync} m_e c^2}{e\delta E}\nonumber\\
&\sim&
26\ {\rm km}
\ \left(\frac{f}{0.1}\right)^{-3/2}\left(\frac{B_*}{10^{12}\ {\rm G}}\right)^{-3/2}\left(\frac{P}{0.1{\rm s}}\right)^{9/2}.
\nonumber
\end{eqnarray}
{As expected, the characteristic synchrotron radiation energy in eq.~(\ref{radgammas}) is independent of $\gamma$ and $\delta B$.}
This acceleration length  is small enough to claim that  accelerated particles indeed reach the synchrotron radiation-reaction limit while they remain inside the midplane of the ECS. The latter calculation, however, may be a gross underestimation of $\gamma_{\rm sync}$ and $l_{\rm acc}$ if the radiating particle pitch angle $\alpha$ is extremely small, as is shown  in \cite{Kalapotharakos2019}. If this is indeed the case and $\alpha\approx 0^\circ$, the accelerated particles will reach the curvature radiation-reaction limit before synchrotron radiation becomes significant.

We can also estimate the total electromagnetic power that is lost over a strip of the ECS with perimeter $\sim 2\pi R_{\rm lc}$ and width $\sim R_{\rm lc}$, which subsequently accelerates particles to their $\gamma$-ray radiation-reaction limit. This estimate yields
\[
L_\gamma \sim ec\delta E\ \sigma\  2\pi R_{\rm lc}^2\approx 2\pi\ fB_{\rm lc}^2 R_{\rm lc}^2c
\]
\begin{equation}
=6.28\times 10^{24}\mbox{erg/s}\ \left(\frac{f}{0.1}\right)^2\left(\frac{B_*}{10^{12}\ {\rm G}}\right)^2\left(\frac{P}{0.1{\rm s}}\right)^{-4}
\end{equation}
\begin{equation}
\sim 0.1\dot{\cal E},
\label{Lgamma}
\end{equation}
where
\[
\dot{\cal E}\sim \frac{B_*^2 r_*^6 \Omega^4}{c^3}\equiv B_{\rm lc}^2 R_{\rm lc}^2 c
\]
\begin{equation}
=5.76\times 10^{25}\mbox{erg/s}\ \left(\frac{B_*}{10^{12}\ {\rm G}}\right)^2\left(\frac{P}{0.1{\rm s}}\right)^{-4}
\end{equation}
is the canonical dipole spindown luminosity \cite{Spitkovsky2006}. 
The relation $L_\gamma\propto \dot{\cal E}$ is rather trivial and is mainly based on the numerically derived result that the accelerating electric field in the midplane of the ECS is proportional to the magnetic field above and below the ECS.  
3PC data suggest that the latter linear relation between $L_\gamma$ and $\dot{\cal E}$ is probably not valid for high $\dot{\cal E}$ \citep[see figure~23 of][]{Smith2023}. 
Since  pulsars emit $\gamma$-rays at the level of $L_\gamma\lsim \dot{\cal E}$, from the above association we deduce that there are not many extra radiating particles in the ECS beyond the minimum number of positrons required to support the electric charge and the electric current in the ECS. 

This simple calculation is consistent with the famous theoretical expression for the $\gamma$-ray pulsar fundamental plane \citep{Kalapotharakos2019}, namely
\begin{equation}
L_\gamma\propto \epsilon_{\rm cut}^{4/3}B_*^{1/6}\dot{\cal E}^{5/12}\ ,
\label{fp}
\end{equation}
where
\begin{equation}
\epsilon_{\rm cut}\equiv (3/2)c\hbar \gamma^3/R_{\rm c}
\label{ecutoff}
\end{equation}
is the cutoff energy of a particle emitting curvature radiation along a trajectory with curvature radius $R_{\rm c}$. If we set $\gamma=\gamma_{\rm curv}$ as obtained above, and $R_{\rm c}=R_{\rm lc}$, eq.~(\ref{fp}) yields $L_\gamma\propto B_*^{2}P^{-4}$, which becomes eq.~(\ref{Lgamma}). Equation~(\ref{fp}) is clearly not satisfied for synchrotron radiation (use eq.~\ref{gammas} instead of eq.~\ref{gammac} in eqs.~\ref{ecutoff} and \ref{fp}). 

We emphasize that eq.~(\ref{fp}) does not teach us anything new about $L_\gamma$. Its importance lies in the fact that this particular combination is satisfied for curvature radiation, but not for synchrotron radiation. This settles the origin of the MeV-GeV pulsar $\gamma$-ray emission as being due to radiation-reaction limited curvature radiation. {The same can be seen directly in the characteristic energies of radiation-reaction limited curvature and synchrotron radiation obtained in eqs.~(\ref{radgammac}) and (\ref{radgammas}), respectively.}

\section{Conclusions}

We showed how far one can go by assuming a steady-state ideal FFE solution with a well-determined steady-state ECS, in addition to which  one can introduce electromagnetic energy dissipation and particle acceleration in the radiation-reaction limit. We were thus able to generate high-energy radiation sky maps with realistic physical parameters without any extrapolations. There exist several other similar examples in the literature in which sky maps were obtained from ideal solutions assuming various radiation mechanisms (e.g., \citet{Bai2010,Kalapotharakos2012} for the pulsar magnetosphere and \citet{DimitropoulosNathanail} for the black hole magnetosphere). In our present work we focused on high-energy radiation produced in the ECS beyond the light cylinder. We argued that the ECS is probably stabilized near the light cylinder by the normal component of the magnetic field $\vv{\delta B}$ because $\delta B/B\sim 0.1\gg S^{-3/4}$, where $S\gg 1$ is the Lundquist number. We also argued that positrons and electrons are probably accelerated by the combined effect of $\vv{\delta E}$ and $\vv{\delta B}$ in its midplane. 
One observational signature that could statistically disentangle between the different radiation prescriptions would be to fit the distribution of radio-lag versus $\gamma$-ray peak-separation as observed by Fermi \citep[3PC;][]{Smith2023}.
Finally, we showed that the simple analytical model of the split-monopole is helpful in approximating high-energy sky maps from numerical simulations, provided its ECS emits only beyond the light cylinder (not from the stellar surface outward).

\begin{acknowledgements}

This work was supported by computational time granted from the National Infrastructure for Research and Technology S.A. (GRNET) in the National HPC facility - ARIS - under project ID pr016005-gpu. It was also supported by the ANR (Agence Nationale de la Recherche) grant number ANR-20-CE31-0010. ID was supported by the Hellenic Foundation for Research and Innovation (HFRI) under the 4th Call for HFRI PhD Fellowships (Fellowship Number: 9239).

\end{acknowledgements}

\section*{Data availability}
The data supporting this article will be shared on reasonable re-
quest to the corresponding author.

\bibliographystyle{aa}
\bibliography{Literature} 

\end{document}